\documentclass[12pt,preprint]{emulateapj}

\usepackage{amsmath}
\usepackage{bm}
\slugcomment{\today}

\shorttitle{Gravitational-Wave Background and Metallicity Evolution}
\shortauthors{Nakazato et al.}

\begin{document}


\title{Gravitational-Wave Background from Binary Mergers and Metallicity Evolution of Galaxies}


\author{Ken'ichiro Nakazato\altaffilmark{1}, Yuu Niino\altaffilmark{2}, and Norichika Sago\altaffilmark{1}}

\email{nakazato@artsci.kyushu-u.ac.jp}


\altaffiltext{1}{Faculty of Arts and Science, Kyushu University, 744 Motooka, Nishi-ku, Fukuoka 819-0395, Japan}
\altaffiltext{2}{Division of Optical \& Infrared Astronomy, National Astronomical Observatory of Japan, 2-21-1 Osawa, Mitaka, Tokyo 181-8588, Japan}


\begin{abstract}
The cosmological evolution of the binary black hole (BH) merger rate and the energy density of the gravitational-wave (GW) background are investigated. To evaluate the redshift dependence of the BH formation rate, BHs are assumed to originate from low-metallicity stars, and the relations between the star formation rate, metallicity and stellar mass of galaxies are combined with the stellar mass function at each redshift. As a result, it is found that when the energy density of the GW background is scaled with the merger rate at the local Universe, the scaling factor does not depend on the critical metallicity for the formation of BHs. Also taking into account the merger of binary neutron stars, a simple formula to express the energy spectrum of the GW background is constructed for the inspiral phase. The relation between the local merger rate and the energy density of the GW background will be examined by future GW observations.
\end{abstract}


%% subject header.  Objects should be in the appropriate "individual"
%% headers (e.g. quasars: individual, stars: individual, etc.) with the
%% additional provision that the total number of headers, including each
%% individual object, not exceed six.  The \objectname{} macro, and its
%% alias \object{}, is used to mark each object.  The macro takes the object
%% name as its primary argument.  This name will appear in the paper
%% and serve as the link's anchor in the electronic edition if the name
%% is recognized by the data centers.  The macro also takes an optional
%% argument in parentheses in cases where the data center identification
%% differs from what is to be printed in the paper.

\keywords{black hole physics --- galaxies: evolution --- gravitational waves --- stars: neutron}


\section{Introduction}\label{sec:intro}

\begin{figure}
\plotone{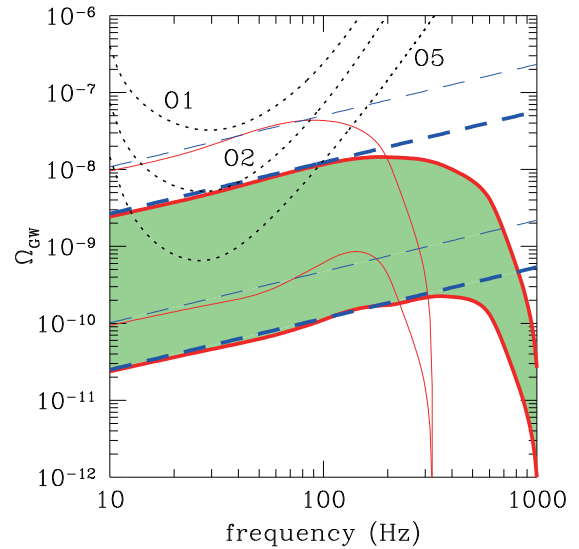}
\epsscale{1.0}
\caption{Energy density spectra of GW background (solid lines) and expected sensitivity of the network of Advanced LIGO and VIRGO (dotted lines). The dashed lines represent the approximation (\ref{eq:gwbgispr}). For solid and dashed lines, the thick ones correspond to the binary BH mass distribution from the results of the Advanced LIGO detectors and the thin ones correspond to the case in which all BH binaries have the same mass as GW150914. The upper and lower ones are for the cases with $(R^{\rm BH}_m(0),t_{\rm min})=(240~{\rm Gpc}^{-3}{\rm yr}^{-1},50~{\rm Myr})$ and $(R^{\rm BH}_m(0),t_{\rm min})=(9~{\rm Gpc}^{-3}{\rm yr}^{-1},5~{\rm Gyr})$, respectively. The sensitivity curves correspond, from top to bottom, to observation runs O1, O2 and O5 from \citet{LIGOc}.}
\label{fig:intro}
\end{figure}

The detection of a gravitational wave (GW) signal by the Advanced LIGO detectors has enabled the use of GWs as a new observational method for astrophysics \citep[][]{LIGOa,LIGOb}. An upgrade over the next several years will increase the sensitivity of Advanced LIGO (Figure~\ref{fig:intro}), and new detectors, KAGRA \citep[][]{aso13} and Advanced VIRGO \citep[][]{virgo}, will soon be installed. Thus, a large amount of observational data on GWs, which are not only signals from individual events but also the background from unresolved events \citep[][]{LIGOc}, will be available in the near future. In any case, we will be able to access unique information on source objects such as black holes (BHs) and neutron stars (NSs). Statistical studies on their binary merger rate will also be possible \citep[][]{LIGOd} when the number of detected merger signals increases. Up to now, two signals (GW150914 and GW151226) and one candidate (LVT151012) have been reported from the first observational run of the Advanced LIGO detectors \citep[][]{LIGOe,LIGOf}.

The first detection of a GW, GW150914, was a merger of two BHs with masses of $36^{+5}_{-4}M_\odot$ and $29^{+4}_{-4}M_\odot$ \citep[][]{LIGOa}. This is interesting because the astrophysical origin of BHs with $\sim$30$M_\odot$, particularly their binaries, is not trivial. Since massive stars with solar abundance (metallicity of $Z_\odot=0.02$) lose a large amount of their mass in the late stage of their evolution, it is difficult for them to form such heavy BHs. In contrast, the mass loss is not efficient for stars with low metallicities. If the mass of the iron core is too high at the end of their life, the stellar core collapse would result in the failure of explosion and the progenitors would completely collapse to a BH \citep[e.g.,][]{fryer99,lieb04,self13a}. Therefore, the BH binaries found as GW150914 are expected to have been formed in a low-metallicity environment \citep[][]{belcz10a,belcz10b,spera15,LIGOb}.

In addition to the above condition, we should consider the formation and evolution process of binary systems of heavy BHs. It is thought that two formation channels are possible \citep[][]{LIGOb}. The first possibility is from isolated binaries of massive stars \citep[][]{tuyu93,kinu14,mdm16,belcz16}. In this scenario, both of the massive stars would result in BH formation while their evolution process would be different from that of single stars. They would experience highly non-conservative mass transfer, common envelope ejection or chemically homogeneous evolution due to strong internal mixing. The second possibility is through dynamical interactions in a dense stellar cluster \citep[][]{pzm00,oleary06,rodrig15,rodrig16}. In this scenario, the binary BH can be formed by a three-body encounter of a single BH with a binary containing another BH.

The precise value of the critical metallicity for the formation of heavy BHs is also an uncertain factor. BH formation may be possible in sub-solar metallicity of $0.5Z_\odot$ or $0.1Z_\odot$ \citep[][]{LIGOb,belcz16}, or may require metal-free (Population III) stars \citep[][]{kinu14}. In any case, the cosmological evolution of metallicity plays a key role in estimating the event rate of binary BH mergers.

In this study, we investigate the cosmological evolution of the binary BH merger rate based on the star formation history of low-metallicity stars. Recently, the metallicity in the high-redshift Universe has been measured by observations of galaxies \citep[e.g.,][]{m08}, as well as the star formation rate (SFR) and stellar mass function \citep[e.g.,][]{da08}. Incidentally, the correlation between the stellar mass, the SFR and the metallicity of galaxies has been studied for various ranges of the cosmic redshift parameter $z$ \citep[e.g.,][]{mannu10,niino12,yabe12,yabe14,zahid14}. Since the BHs are formed as remnants of short-lived massive stars, we combine the SFR with the metallicity of individual galaxies, which was not considered in the previous study by \citet{LIGOc}.

This paper is organized as follows. In \S~\ref{sec:csfrd}, we introduce the model of galaxy evolution and apply it to the formation history of low-metallicity stars. We also examine the model adopted in \citet{LIGOc}. In \S~\ref{sec:mrgwbg}, we present the formulation of the binary BH merger rate. We also investigate the GW background from binary BH mergers and derive the relation between the merger rate at the local Universe and the energy density of the GW background. Furthermore, we also consider binary NS mergers and the GW background from them. Finally, \S~\ref{sec:disc} is devoted to discussion.

\section{Cosmic Star Formation Rate Density of Low-Metallicity Stars}\label{sec:csfrd}

Heavy BHs with mass $\sim$30$M_\odot$ are expected to be formed in a low-metallicity environment below a critical metallicity, $Z_{\rm crit}$. Therefore the BH formation rate should be proportional to the cosmic star formation rate density (CSFRD) of low-metallicity stars. We investigate the dependence on $Z_{\rm crit}$ considering the CSFRD of stars with metallicity below $Z_{\rm crit}$, as done in \citet{LIGOc}. Note that, while \citet{LIGOb,LIGOc} assumed $Z_{\rm crit}=0.5 Z_\odot$ for their fiducial model, \citet{belcz16} showed that the formation of the binary heavy BHs requires $< \! 0.1Z_\odot$. In this section, we first describe our standard model based on the observational data of galaxies. Next, for comparison, we consider the alternative model for the CSFRD of low-metallicity stars following \citet{LIGOc}.

\subsection{Models of Star Formation Rate and Metallicity Evolution of Galaxies}\label{sec:sfrme}

To derive the CSFRD of low-metallicity stars, we consider the stellar mass ($M_\ast$), SFR ($\dot M_\ast$) and metallicity [12+log$_{10}$(O/H)] of galaxies. Here, SFR is the total mass of stars formed in the galaxy per unit time. For our standard model, we adopt the redshift evolutions of the galaxy stellar mass function and the mass-dependent SFR proposed by \citet{da08} and the redshift-dependent mass-metallicity relation from \citet{m08}. These models were also utilized to evaluate a spectrum of supernova relic neutrinos in a previous study \citep{self15}.

Since, at redshift $z$, $\dot M_\ast(M_\ast,z)$ is SFR of a galaxy with a stellar mass of $M_\ast$ and $\phi_{\rm SMF}(M_\ast,z){\rm d}M_\ast$ is a number density of galaxies within a stellar mass bin of $[M_\ast,M_\ast+{\rm d}M_\ast]$, the total CSFRD is given by
\begin{equation}
\dot \rho_\ast(z) = \int^\infty_0 \dot M_\ast(M_\ast,z) \phi_{\rm SMF}(M_\ast,z) \ {\rm d}M_\ast ,
\label{eq:csfrd}
\end{equation}
with the stellar mass function $\phi_{\rm SMF}(M_\ast,z)$. In \citet{da08}, the stellar mass function is assumed to have a Schechter form,
\begin{eqnarray}
\phi_{\rm SMF}(M_\ast,z) \ {\rm d}M_\ast & = & \phi_0(z)\left(\frac{M_\ast}{M_0^{\rm DA08}(z)}\right)^{-1.3} \nonumber \\
 & \times & \exp\left(-\frac{M_\ast}{M_0^{\rm DA08}(z)}\right)\frac{{\rm d}M_\ast}{M_0^{\rm DA08}(z)},
\label{eq:smf}
\end{eqnarray}
with the best-fitting parameterization
\begin{subequations}
\begin{equation}
\phi_0(z)=0.0031\times(1+z)^{-1.07} \ {\rm Mpc}^{-3} \ {\rm dex}^{-1},
\label{eq:smf-phi0}
\end{equation}
\begin{equation}
\log_{10}\left(\frac{M_0^{\rm DA08}(z)}{M_\odot}\right)=11.35-0.22\times\ln(1+z).
\label{eq:smf-m0}
\end{equation}
\label{eq:smf-spl}
\end{subequations}
As expressed in Equation~(\ref{eq:smf}), the stellar mass function has a sharp exponential cutoff above $M_0^{\rm DA08}$. The SFR is a function of the stellar mass and redshift and is written as \citep[][]{da08}
\begin{equation}
\dot M_\ast(M_\ast,z) = \dot M^0_\ast(z) \left(\frac{M_\ast}{M_1^{\rm DA08}(z)}\right)^{0.6} \exp\left(-\frac{M_\ast}{M_1^{\rm DA08}(z)}\right),
\label{eq:sfr}
\end{equation}
with
\begin{subequations}
\begin{equation}
\dot M^0_\ast(z) = 1.183 \times (1+z)^{5.5} \exp(-0.78z) \ M_\odot \ {\rm yr}^{-1},
\label{eq:sfr-mdot0}
\end{equation}
\begin{equation}
M_1^{\rm DA08}(z) = 2.7 \times 10^{10} \times (1+z)^{2.1} \ M_\odot.
\label{eq:sfr-m1}
\end{equation}
\label{eq:sfr-spl}
\end{subequations}
Note that the analytic form of Equation~(\ref{eq:sfr-mdot0}) was determined to fit not only the original data of galaxies observed by \citet{drory05} but also the CSFRD at the local universe, $\dot \rho_\ast(0)=0.02M_\odot~{\rm yr}^{-1}~{\rm Mpc}^{-3}$ \citep{self15}. Roughly speaking, the SFR is higher for galaxies with higher stellar mass, while the specific SFR, $\dot M_\ast/M_\ast$, is higher for galaxies with lower stellar mass. Nevertheless, star formation is strongly suppressed for galaxies with enough high stellar mass and $M_1^{\rm DA08}$ corresponds to the mass above which SFR begins to decline.

\begin{figure*}
\plotone{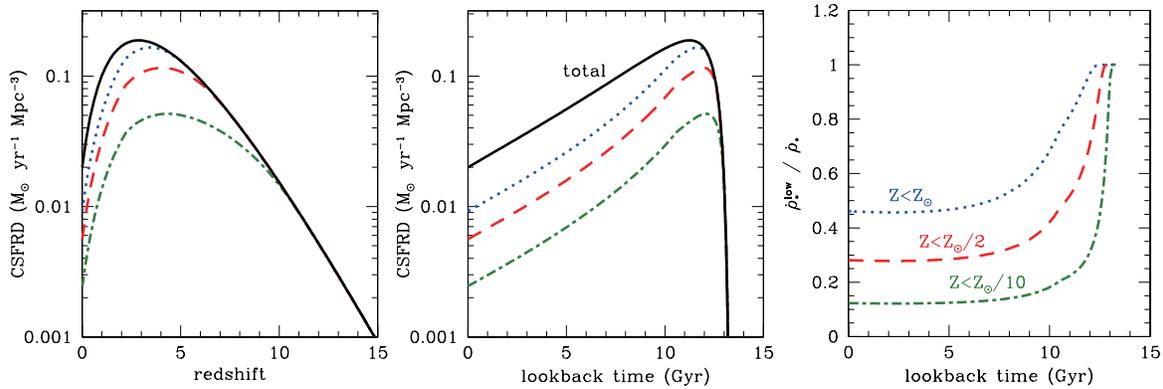}
\epsscale{1.0}
\caption{Total CSFRD and CSFRD of stars with metallicity below $Z_{\rm crit}$ as a function of redshift (left) and lookback time (center), and fraction of stars formed with metallicity below $Z_{\rm crit}$ as a function of lookback time (right) for our standard model. The solid lines represent the total CSFRD in the left and central panels. The other lines correspond, from bottom to top, to $Z_{\rm crit}=0.1Z_\odot$, $0.5Z_\odot$ and $Z_\odot$.}
\label{fig:csfrd}
\end{figure*}

The CSFRD for a metallicity below $Z_{\rm crit}$ is given by
\begin{equation}
\dot \rho^{\rm low}_\ast(z, Z_{\rm crit}) = \displaystyle \int^{M_\ast(z,Z_{\rm crit})}_0 \dot M_\ast(M_\ast^\prime,z) \phi_{\rm SMF}(M_\ast^\prime,z) \ {\rm d}M_\ast^\prime,
\label{eq:zcsfrd}
\end{equation}
where $M_\ast(z,Z)$ is the stellar mass of a galaxy with metallicity $Z$ at redshift $z$. It is known that lower-metallicity galaxies have systematically lower stellar mass. In \citet{m08}, the galaxy mass-metallicity relation was expressed as 
\begin{eqnarray}
\log_{10}\left(\frac{Z}{Z_\odot}\right) +8.69 & = & -0.0864 \times (\log_{10}M_\ast -\log_{10}M_0^{\rm M08})^2 \nonumber \\
 & & + K_0^{\rm M08},
\label{eq:M08}
\end{eqnarray}
where we adopt the best fit values of $\log_{10}M_0^{\rm M08}$ and $K_0^{\rm M08}$ at different redshifts from case {\it a} in Table~5 of \citet{m08}. This relation is applicable for stellar masses of $8.5 \le \log_{10}(M_\ast/M_\odot) \le 11.2$, which correspond to $8.2 \le 12+\log_{10}({\rm O}/{\rm H}) \le 9$ at $z=0.7$ and $7.2 \le 12+\log_{10}({\rm O}/{\rm H}) \le 8.6$ at $z=3.5$. Note that the solar metallicity ($Z_\odot=0.02$) is assumed to correspond to the oxygen abundance of $12+\log_{10}({\rm O}/{\rm H}) = 8.69$ \citep[][]{alle01}, i.e.,
\begin{equation}
\log_{10}\left(\frac{Z}{Z_\odot}\right) = 12+\log_{10}({\rm O}/{\rm H}) -8.69.
\label{eq:solmet}
\end{equation}
Using Equation (\ref{eq:M08}), we calculate $M_\ast(z,Z)$ and, thereby, $\dot \rho^{\rm low}_\ast(z, Z_{\rm crit})$.

We show the total CSFRD and CSFRD of low-metallicity stars in Figure~\ref{fig:csfrd}. The resultant total CSFRD is lower than that in \citet{hb06}, which has often been cited. In contrast, the theoretical models predict a much lower CSFRD \citep[e.g.,][]{koba13} and our CSFRD model lies between them for $z \lesssim 2$.  Incidentally, in the redshift range of $0 \le z \le 2$, the difference from the recent model of \citet{md14} is not large; our CSFRD model is 10--30\% higher. The total CSFRD has a peak near the redshift $z=3$, or equivalently the lookback time of 11~Gyr, and declines towards the present epoch. In Figure~\ref{fig:csfrd}, we also show the fraction of stars formed with metallicity below $Z_{\rm crit}$, $\dot \rho^{\rm low}_\ast(z, Z_{\rm crit})/\dot \rho_\ast(z)$, as a function of the lookback time. For $Z_{\rm crit} \le 0.5Z_\odot$, the fraction has not varied much over the last 8~Gyr (i.e., redshift $z \lesssim 1$), while the total CSFRD is decreasing. This trend originates from the models of galaxy evolution adopted in this study. Firstly, the stellar mass function of \citet{da08} does not evolve significantly in this period. Secondly, when the SFR is drawn as a function of the stellar mass, the slope in the low-mass range does not depend on the redshift, and the peak mass, $M_1^{\rm DA08}$, becomes higher at a high redshift \citep[see the top panel of Figure~3 in][]{da08}. On the other hand, in \citet{m08}, the galaxy mass-metallicity relation moves toward higher masses but its shape is preserved at a high redshift, which is also described in \citet{sava05}. As a result, these two shifts balance out and the fraction of stars formed in a low-metallicity environment is almost constant for $z \lesssim 1$.

\subsection{Alternative Model for Fraction of Low-Metallicity Stars}\label{sec:alter}

Here, we construct the alternative model for the CSFRD of low-metallicity stars following \citet{LIGOc}, which was also adopted in \citet{call16}. It is based on the mean metallicity at the redshift $z$, which is written as
\begin{equation}
 Z_{\rm mean}(z) = Z_{\rm mean}(0) \ \frac{\int_z^{z_{\rm max}} \frac{\dot \rho_\ast(z^\prime) \ {\rm d}z^\prime}{\bigl\{ H_0(1+z^\prime)\sqrt{\Omega_m (1+z^\prime)^3 + \Omega_\Lambda} \bigr\}}}{\int_0^{z_{\rm max}} \frac{\dot \rho_\ast(z^\prime) \ {\rm d}z^\prime}{\bigl\{ H_0(1+z^\prime)\sqrt{\Omega_m (1+z^\prime)^3 + \Omega_\Lambda} \bigr\}}},
\label{eq:solmet}
\end{equation}
with cosmological constants $H_0=70~{\rm km}~{\rm s}^{-1} \,{\rm Mpc}^{-1}$, $\Omega_m=0.3$ and $\Omega_\Lambda=0.7$ \citep[][]{md14}. We set $z_{\rm max}=20$ \citep[][]{belcz16}. For the mean metallicity at $z=0$, we adopt the value of $Z_{\rm mean}(0) =0.5 Z_\odot$ \citep[][]{vang15}. Note that \citet{LIGOc} adopted the mean metallicity-redshift relation of \citet{md14} but rescaled it to account for local observations \citep[][]{vang15,belcz16}. For convenience of comparison, we use the same function for the total CSFRD, $\dot \rho_\ast(z)$, as that derived in \S~\ref{sec:sfrme}. At each redshift, the metallicity is assumed to have a log-normal distribution with a standard deviation of 0.5~dex around the mean. Incidentally, this is the metallicity dispersion of the interstellar medium measured for damped Ly $\alpha$ absorbers \citep[][]{dvor15}, and the metallicity evolution model based on damped Ly $\alpha$ data was considered by \citet{dvor16} to calculate the merger rate of binary BHs.

\begin{figure*}
\plotone{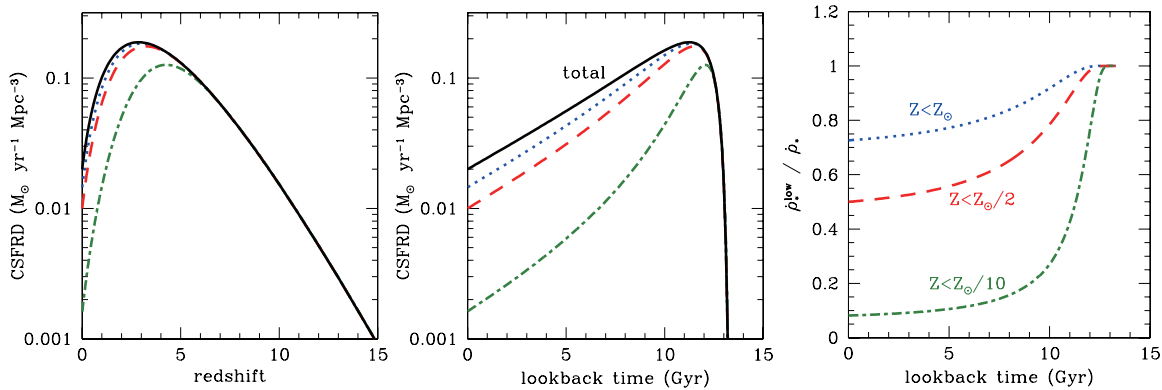}
\epsscale{1.0}
\caption{Same as Figure~\ref{fig:csfrd} but for the alternative model given in \S~\ref{sec:alter}.}
\label{fig:csfrda}
\end{figure*}

In Figure~\ref{fig:csfrda}, we show the CSFRD and the fraction of low-metallicity stars for the alternative model. In this model, the formation of metal-poor stars is reduced for lookback times within 10~Gyr (i.e., redshift $z \lesssim 2$) compared with our standard model. Before closing this section, we emphasize that Equation~(\ref{eq:solmet}) gives the mean metallicity of the Universe and does not purely reflect the metallicity of star-forming regions. Since short-lived massive stars are responsible for the formation of BHs and NSs, our standard model, in which the CSFRD of low-metallicity stars is considered with the stellar mass function and SFR of galaxies, is more reasonable for the purpose of this study.

\section{Merger Rate and Gravitational-Wave Background}\label{sec:mrgwbg}

In this section, we investigate the merger rate of binary BHs formed in a low-metallicity environment. The merger rate is calculated as a function of redshift utilizing the models of metallicity evolution in \S~\ref{sec:csfrd}. For this purpose, the delay time between the binary BH formation and the merger is needed. In particular, we investigate the dependence on the minimum delay time, while \citet{LIGOc} assumed $t_{\rm min}=50$~Myr for their fiducial model. Using the derived merger rate, we study the contribution of binary BH mergers to the GW background. Here, we focus on the relation between the local merger rate and the energy density of the GW background. Furthermore, we consider the contribution of binary NSs. While we again utilize the models of metallicity evolution in \S~\ref{sec:csfrd}, NSs are formed in both low- and high-metallicity environments. Therefore, we take into account the difference in the formation rate between low- and high-metallicity environments.

\subsection{Cosmological Evolution of Merger Rate}\label{sec:mgrate}

We consider the merger rate of binary BHs originating from a low-metallicity environment with metallicity below $Z_{\rm crit}$ as in \citet{LIGOc}. It is determined by convolving the binary BH formation rate $R^{\rm BH}_f(z)$ with the delay time distribution $P(t_d)$ \citep[e.g.,][]{naker07,LIGOc}:
\begin{equation}
R^{\rm BH}_m(z_m) = \int^{t_{\rm max}}_{t_{\rm min}} R^{\rm BH}_f(z_f) P(t_d) \ {\rm d} t_d,
\label{eq:mgrate}
\end{equation}
where the redshift at the formation time $z_f$ is related to the redshift at the merger time $z_m$ and delay time $t_d$ as $T(z_f)=T(z_m)+t_d$, denoting the lookback time at redshift $z$ as $T(z)$. We assume that the delay time distribution follows $P(t_d) \propto 1/t_d$ for $t_d > t_{\rm min}$ \citep[][]{LIGOc,belcz16}, where $t_{\rm min}$ is the minimum delay time for a binary BH to evolve until merger. The maximum delay time $t_{\rm max}$ is set to the Hubble time. Note that delay time distribution of $\propto \! 1/t_d$ is usually assumed for isolated binaries. Incidentally the merger rate of binaries formed dynamically in clusters is inversely proportional to the age of the cluster after the depletion of binary formation \citep[][]{oleary06}.

The binary BH formation rate is assumed to be proportional to the CSFRD of low-metallicity stars below $Z_{\rm crit}$ as
\begin{equation}
R^{\rm BH}_f(z) = \zeta_{\rm BH} \ \dot \rho^{\rm low}_\ast(z, Z_{\rm crit}),
\label{eq:fmrate}
\end{equation}
with conversion coefficient $\zeta_{\rm BH}$. While some uncertainties regarding the formation process of a binary BH, such as the binary formation rate, binary evolution model and BH formation rate, are encapsulated into $\zeta_{\rm BH}$, in this paper we do not evaluate $\zeta_{\rm BH}$ explicitly. In other words, we focus on the physics free from these uncertainties.

\begin{figure*}
\plotone{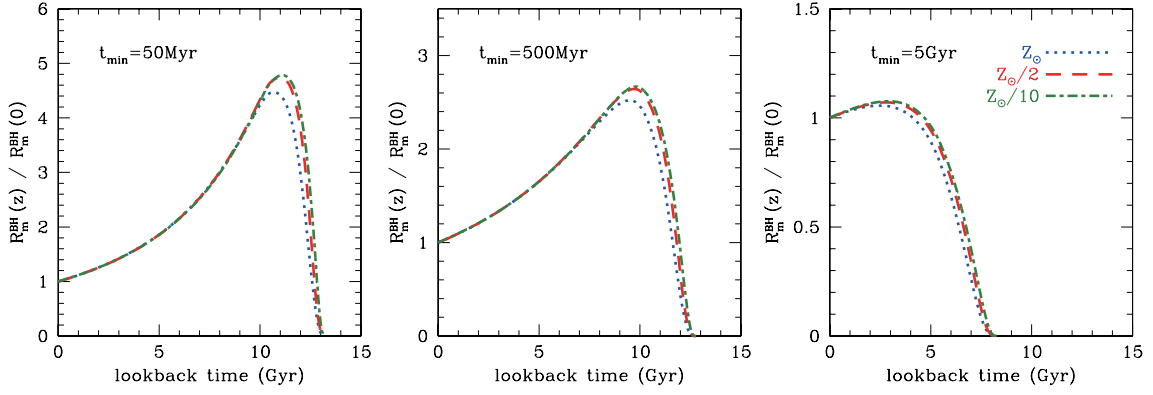}
\epsscale{1.0}
\caption{Scaled merger rate of binary BHs as a function of lookback time for our standard model. The values at the local Universe are set to unity. The left, central and right panels correspond to the models with the minimum delay time $t_{\rm min}=50$~Myr, 500~Myr and 5~Gyr, respectively. The dot-dashed, dashed and dotted lines correspond to the critical metallicities $Z_{\rm crit}=0.1Z_\odot$, $0.5Z_\odot$ and $Z_\odot$, respectively.}
\label{fig:mgrate}
\end{figure*}

In Figure~\ref{fig:mgrate}, the merger rates of binary BHs for our standard model are plotted as a function of the lookback time. They are shown as a ratio to the values at the local Universe ($z=0$) to be independent of the conversion coefficient $\zeta_{\rm BH}$, which is not evaluated. We can see that the evolution of the merger rate depends on the minimum delay time $t_{\rm min}$; the rise time of the merger rate becomes later and the ratio of the peak to local merger rates decreases for the models with a longer delay time. On the other hand, the scaled merger rate is insensitive to the critical metallicity, especially for $Z_{\rm crit} \le 0.5Z_\odot$. This is because, for redshift $z \lesssim 1$, the fraction of stars formed below $Z_{\rm crit}$ is almost time independent, and hence the evolution of the scaled formation rate of low-metallicity stars is insensitive to $Z_{\rm crit}$. 

\begin{figure*}
\plotone{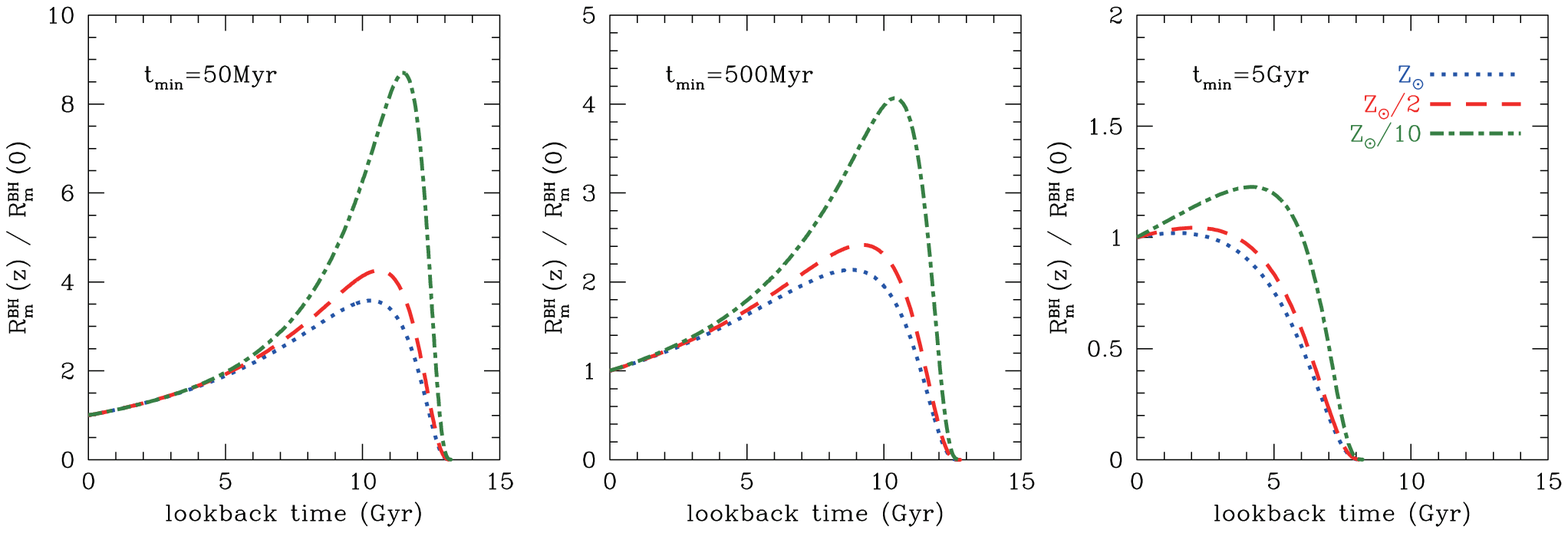}
\epsscale{1.0}
\caption{Same as Figure~\ref{fig:mgrate} but for the alternative model given in \S~\ref{sec:alter}.}
\label{fig:mgratea}
\end{figure*}

The merger rates for the alternative model are shown in Figure~\ref{fig:mgratea}. In this case, the merger rate varies with $Z_{\rm crit}$. In particular, for a case with a lower critical metallicity, binary BHs are formed mainly at a high redshift and the evolution of the merger rate follows the delay time distribution. Therefore, the ratio of the peak to local merger rates is larger for the models with a lower $Z_{\rm crit}$ and/or a shorter delay time.

~\\

\subsection{Gravitational-Wave Background from Binary Inspirals}\label{sec:gwbg}

The GWs emitted from the orbital motion of merging binaries create a GW background. Its energy density spectrum is given by \citep[][]{LIGOc}
\begin{eqnarray}
\Omega^{\rm BH}_{\rm GW}(f) = \frac{f}{\rho_c}\int_0^{z_{\rm max}} {\rm d}z \Biggl[ & & \frac{R^{\rm BH}_m(z)}{H_0(1+z)\sqrt{\Omega_m (1+z)^3 + \Omega_\Lambda}} \nonumber \\
 & & \times \frac{{\rm d}E^{\rm BH}_{\rm GW}(f^\prime)}{{\rm d} f^\prime} \Biggr],
\label{eq:gwbg}
\end{eqnarray}
where the critical energy density is given by $\rho_c=3H_0^2c^2/8\pi G$ with velocity of light $c$ and gravitational constant $G$ \citep[see also][]{zhu11,wu12}. The frequency on the Earth, $f$, is related to that at redshift $z$, $f^\prime$, as $f^\prime=(1+z)f$. The spectral energy density ${\rm d}E^{\rm BH}_{\rm GW}(f)/{\rm d} f$ originates from each merging binary. Here, bearing in mind the sensitive frequency band of detectors (10--50~Hz), we use the following approximation for the spectral energy density: 
\begin{equation}
\frac{{\rm d}E^{\rm BH}_{\rm GW}(f)}{{\rm d} f} \simeq \frac{(G\pi)^{2/3}}{3} \ M_{c,{\rm BH}}^{5/3}f^{-1/3},
\label{eq:egwisp}
\end{equation}
where $M_{c,{\rm BH}}$ is the chirp mass of the binary. This formula closely approximates the spectrum below $\sim$100~Hz, where the contribution from the inspiral phase is dominant. Substituting Equation (\ref{eq:egwisp}) into Equation (\ref{eq:gwbg}), we obtain
\begin{eqnarray}
\Omega^{\rm BH}_{\rm GW}(f) & \simeq & \frac{(G\pi)^{2/3}}{3\rho_c H_0} \ M_{c,{\rm BH}}^{5/3} \ R^{\rm BH}_m(0) \nonumber \\
 & \times & \left\{\int_0^{z_{\rm max}} {\rm d}z \frac{R^{\rm BH}_m(z)/R^{\rm BH}_m(0)}{(1+z)^{4/3}\sqrt{\Omega_m (1+z)^3 + \Omega_\Lambda}} \right\} \nonumber \\
 & \times & f^{2/3}.
\label{eq:gwbgisp}
\end{eqnarray}
Note that, to compute the integral over redshift $z$, the merger rate is scaled with the local value because the conversion coefficient $\zeta_{\rm BH}$ is not evaluated again. 

\begin{figure}
\plotone{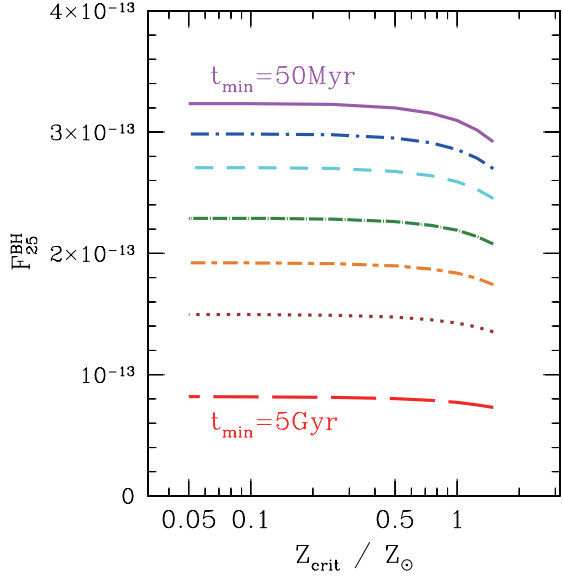}
\epsscale{1.0}
\caption{Critical metallicity, $Z_{\rm crit}$, dependence of the factor $F^{\rm BH}_{25}$ defined in Equation (\ref{eq:gwbgispr}). The lines correspond, from top to bottom, to the cases with minimum delay times $t_{\rm min}=50$~Myr, 100~Myr, 200~Myr, 500~Myr, 1~Gyr, 2~Gyr and 5~Gyr.}
\label{fig:f25bh}
\end{figure}

\begin{figure}
\plotone{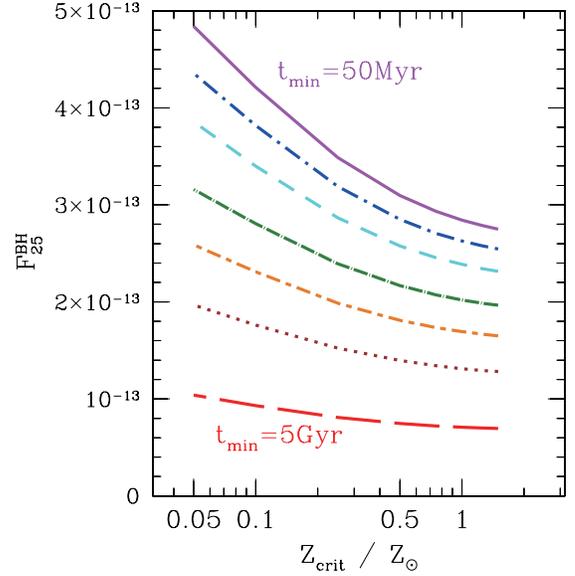}
\epsscale{1.0}
\caption{Same as Figure~\ref{fig:f25bh} but for the alternative model given in \S~\ref{sec:alter}.}
\label{fig:f25bha}
\end{figure}

We rewrite Equation (\ref{eq:gwbgisp}) as
\begin{equation}
\Omega^{\rm BH}_{\rm GW}(f) \simeq F^{\rm BH}_{25} \left( \frac{M_{c,{\rm BH}}}{M_\odot} \right)^{5/3} \left( \frac{R^{\rm BH}_m(0)}{{\rm Gpc}^{-3}{\rm yr}^{-1}} \right) \left( \frac{f}{25~{\rm Hz}} \right)^{2/3},
\label{eq:gwbgispr}
\end{equation}
and determine the factor $F^{\rm BH}_{25}$. As is the case for the merger rate, the critical metallicity $Z_{\rm crit}$ and the minimum delay time $t_{\rm min}$ are needed to determine $F^{\rm BH}_{25}$. For our standard model, we show their dependences of $F^{\rm BH}_{25}$ in Figure~\ref{fig:f25bh}. Since the scaled merger rate does not depend on $Z_{\rm crit}$, $F^{\rm BH}_{25}$ also does not depend on $Z_{\rm crit}$ for $Z_{\rm crit} \le 0.5Z_\odot$. In contrast, for increasing $t_{\rm min}$, the value of $F^{\rm BH}_{25}$ decreases or, equivalently, the energy density of the GW background becomes lower. This is because the merger rate of the past Universe relative to the local Universe decreases for the models with a longer delay time. Furthermore, we find that, for $Z_{\rm crit} \le 0.5Z_\odot$, the minimum delay time dependence of $F^{\rm BH}_{25}$ can be fitted by
\begin{eqnarray}
F^{\rm BH}_{25} & = & c^{\rm BH}_0 + c^{\rm BH}_1 \log_{10}\left( \frac{t_{\rm min}}{50~{\rm Myr}} \right) \nonumber \\
 & & + c^{\rm BH}_2 \left\{ \log_{10}\left( \frac{t_{\rm min}}{50~{\rm Myr}} \right) \right\}^2,
\label{eq:f25bh}
\end{eqnarray}
with $c^{\rm BH}_0=3.2\times 10^{-13}$, $c^{\rm BH}_1=-6.6\times 10^{-14}$ and $c^{\rm BH}_2=-2.6\times 10^{-14}$. Note that, with $M_{c,{\rm BH}}=28M_\odot$, $R^{\rm BH}_m(0)=16^{+38}_{-13}$~Gpc$^{-3}$yr$^{-1}$, $Z_{\rm crit}=0.5Z_\odot$ and $t_{\rm min}=50~{\rm Myr}$, our standard model gives $\Omega^{\rm BH}_{\rm GW}(f=25~{\rm Hz})=1.3^{+3.2}_{-1.1}\times 10^{-9}$, which is close to the value in \citet{LIGOc} with the same input parameters. In contrast, as shown in Figure~\ref{fig:f25bha}, $F^{\rm BH}_{25}$ is a function of not only $t_{\rm min}$ but also $Z_{\rm crit}$ for the alternative model. Incidentally, the alternative model gives $\Omega^{\rm BH}_{\rm GW}(f=25~{\rm Hz})=1.3^{+3.0}_{-1.1}\times 10^{-9}$ with the above inputs.

\subsection{Contribution of Binary Neutron Stars}\label{sec:bns}

All known binary NSs are systems that contain at least one radio pulsar and they are also the target of GW astronomy \citep[e.g.,][]{hultay,kim15}. Here, we consider their merger rate and contribution to the GW background in our framework of metallicity evolution. Since binary NSs are formed in both low- and high-metallicity environments, we write their formation rate as
\begin{eqnarray}
R^{\rm NS}_f(z) & = & \zeta_{\rm NS}^{\rm lowZ} \ \dot \rho^{\rm low}_\ast(z, Z_{\rm crit}) \nonumber \\
 & & + \zeta_{\rm NS}^{\rm highZ} \left\{ \dot \rho_\ast(z) - \dot \rho^{\rm low}_\ast(z, Z_{\rm crit}) \right\},
\label{eq:fmratens}
\end{eqnarray}
with the conversion coefficients $\zeta_{\rm NS}^{\rm lowZ}$ and $\zeta_{\rm NS}^{\rm highZ}$. When we assume that some massive stars become not NSs but heavy BHs in a low-metallicity environment, the conversion coefficient in a high-metallicity environment is larger than that in a low-metallicity environment, i.e., $\zeta_{\rm NS}^{\rm lowZ}\le\zeta_{\rm NS}^{\rm highZ}$. Therefore, they are related by a parameter $x$ ($0\le x\le 1$) as
\begin{equation}
\zeta_{\rm NS}^{\rm lowZ} = x \ \zeta_{\rm NS}^{\rm highZ},
\label{eq:defx}
\end{equation}
and we obtain
\begin{equation}
R^{\rm NS}_f(z) = \zeta_{\rm NS}^{\rm highZ} \left\{ \dot \rho_\ast(z) +(x-1) \ \dot \rho^{\rm low}_\ast(z, Z_{\rm crit}) \right\}.
\label{eq:fmratens2}
\end{equation}
If binary NSs are the main population rather than binary BHs even in a low-metallicity environment, the parameter $x$ is close to 1. Note that the formation rate of binary NSs is proportional to $\zeta_{\rm NS}^{\rm highZ}$, which is not investigated explicitly in this study.

\begin{figure*}
\plotone{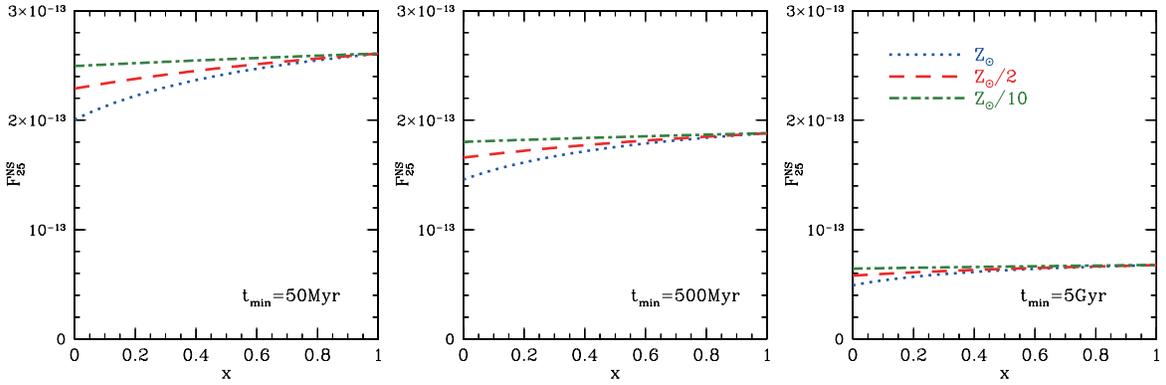}
\epsscale{1.0}
\caption{Factor $F^{\rm NS}_{25}$ defined in Equation (\ref{eq:gwbgisprns}) as a function of parameter $x$ defined in Equation (\ref{eq:defx}). The left, central and right panels correspond to the models with the minimum delay time $t_{\rm min}=50$~Myr, 500~Myr and 5~Gyr, respectively. The dot-dashed, dashed and dotted lines correspond to the critical metallicities $Z_{\rm crit}=0.1Z_\odot$, $0.5Z_\odot$ and $Z_\odot$, respectively.}
\label{fig:f25ns}
\end{figure*}

\begin{figure*}
\plotone{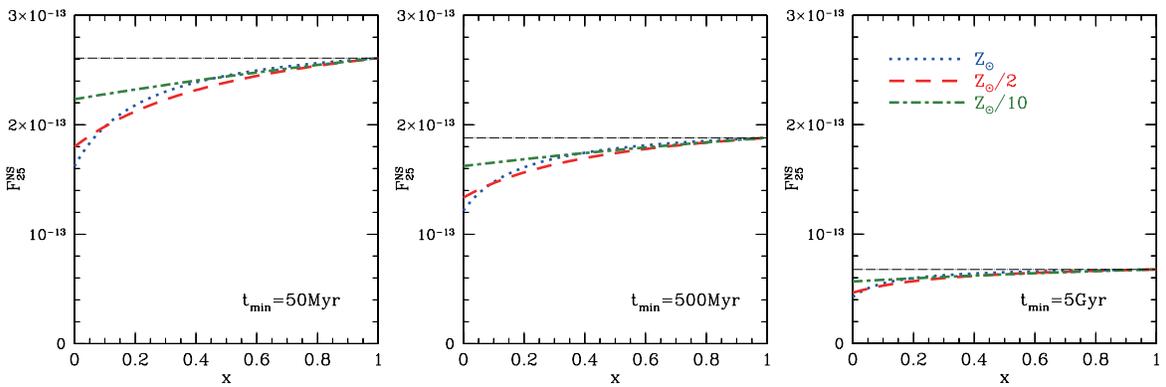}
\epsscale{1.0}
\caption{Same as Figure~\ref{fig:f25ns} but for the alternative model given in \S~\ref{sec:alter}. Thin horizontal lines represent the limit as $Z_{\rm crit} \to 0$.}
\label{fig:f25nsa}
\end{figure*}

The merger rate of binary NSs is given by
\begin{equation}
R^{\rm NS}_m(z_m) = \int^{t_{\rm max}}_{t_{\rm min}} R^{\rm NS}_f(z_f) P(t_d) \ {\rm d} t_d,
\label{eq:mgratens}
\end{equation}
where the meanings of $z_f$, $z_m$ and $P(t_d)$ are the same as in Equation (\ref{eq:mgrate}). Furthermore, the energy density spectrum of the GW background from binary NSs is given by
\begin{eqnarray}
\Omega^{\rm NS}_{\rm GW}(f) = \frac{f}{\rho_c}\int_0^{z_{\rm max}} {\rm d}z \Biggl[ & & \frac{R^{\rm NS}_m(z)}{H_0(1+z)\sqrt{\Omega_m (1+z)^3 + \Omega_\Lambda}} \nonumber \\
 & & \times \frac{{\rm d}E^{\rm NS}_{\rm GW}(f^\prime)}{{\rm d} f^\prime} \Biggr].
\label{eq:gwbgns}
\end{eqnarray}
For the spectral energy density ${\rm d}E^{\rm NS}_{\rm GW}(f)/{\rm d} f$, we again assume the approximation of the inspiral phase with a chirp mass $M_{c,{\rm NS}}$, and we rewrite Equation (\ref{eq:gwbgns}) as
\begin{equation}
\Omega^{\rm NS}_{\rm GW}(f) \simeq F^{\rm NS}_{25} \left( \frac{M_{c,{\rm NS}}}{M_\odot} \right)^{5/3} \left( \frac{R^{\rm NS}_m(0)}{{\rm Gpc}^{-3}{\rm yr}^{-1}} \right) \left( \frac{f}{25~{\rm Hz}} \right)^{2/3},
\label{eq:gwbgisprns}
\end{equation}
with a factor $F^{\rm NS}_{25}$. To determine $F^{\rm NS}_{25}$, we need not only the critical metallicity $Z_{\rm crit}$ and the minimum delay time $t_{\rm min}$ but also the parameter $x$. Their dependences of $F^{\rm NS}_{25}$ for our standard model are shown in Figure~\ref{fig:f25ns}. For the case with $Z_{\rm crit} \le 0.5Z_\odot$ and $0.5 \le x \le 1$, $F^{\rm NS}_{25}$ does not depend on $Z_{\rm crit}$ and $x$ because the term associated with low-metallicity stars, $(x-1)\dot \rho^{\rm low}_\ast(z, Z_{\rm crit})$, is small compared with the total CSFRD, $\dot \rho_\ast(z)$, in Equation (\ref{eq:fmratens2}). As in Equation (\ref{eq:f25bh}), for $Z_{\rm crit} \le 0.5Z_\odot$ and $0.5 \le x \le 1$, the $t_{\rm min}$ dependence of $F^{\rm BH}_{25}$ is fitted by
\begin{eqnarray}
F^{\rm NS}_{25} & = & c^{\rm NS}_0 + c^{\rm NS}_1 \log_{10}\left( \frac{t_{\rm min}}{50~{\rm Myr}} \right) \nonumber \\
 & & + c^{\rm NS}_2 \left\{ \log_{10}\left( \frac{t_{\rm min}}{50~{\rm Myr}} \right) \right\}^2,
\label{eq:f25ns}
\end{eqnarray}
with $c^{\rm NS}_0=2.5\times 10^{-13}$, $c^{\rm NS}_1=-4.5\times 10^{-14}$ and $c^{\rm NS}_2=-2.4\times 10^{-14}$.

In Figure~\ref{fig:f25nsa}, we show the factor $F^{\rm NS}_{25}$ for the alternative model. Also for this model, as $Z_{\rm crit} \to 0$ and/or $x \to 1$, the value of $F^{\rm NS}_{25}$ converges to the same limit as our standard model given in Equation~(\ref{eq:f25ns}). This is because the same model for the total CSFRD is adopted in both cases and the merger rate of binary NSs is mainly determined by the total CSFRD in this limit. Therefore, if the critical metallicity is sufficiently low and/or the parameter $x$ is close to 1, the factor $F^{\rm NS}_{25}$ does not depend on the metallicity distribution of star formation.

\section{Discussion}\label{sec:disc}

In this section, we discuss the implications of this study for future GW astronomy. In the following, we only take into account the standard model for the CSFRD of low-metallicity stars described in \S~\ref{sec:sfrme}.

In the previous sections, we used the approximation for the GW spectrum in the inspiral phase given in Equation~(\ref{eq:egwisp}). In actuality, the spectrum has a cutoff at the high-frequency end, which roughly corresponds to the frequency at the merger. The cutoff frequency is higher for lower-mass mergers. Here, to study the validity of this approximation, we utilize the spectrum of binary BH merger proposed by \citet{ajith11} for comparison. For illustration, we adopt the critical metallicity $Z_{\rm crit}=Z_\odot/\sqrt{5}$ from the heavy-BH formation model in \citet{self13a,self15}, while the choice of $Z_{\rm crit}$ does not affect the result for the GW spectrum. Furthermore, we assume the distribution of binary chirp mass based on the results of the first observational run of the Advanced LIGO detectors \citep[][]{LIGOf}. The event based merger rates are evaluated to be $R_{\rm GW150914}=3.4~{\rm Gpc}^{-3}{\rm yr}^{-1}$, $R_{\rm LVT151012}=9.4~{\rm Gpc}^{-3}{\rm yr}^{-1}$ and $R_{\rm GW151226}=37~{\rm Gpc}^{-3}{\rm yr}^{-1}$ for chirp masses of $M_{c,{\rm GW150914}}=28.1M_\odot$, $M_{c,{\rm LVT151012}}=15.1M_\odot$ and $M_{c,{\rm GW151226}}=8.9M_\odot$, respectively. Hereafter, fixing the ratio of these rates, we consider the dependence on the total merger rate at the local Universe, $R^{\rm BH}_m(0)$. For simplicity, we ignore the spins of BHs, or equivalently, the effective spin parameter is set to zero.

\begin{figure*}
\plotone{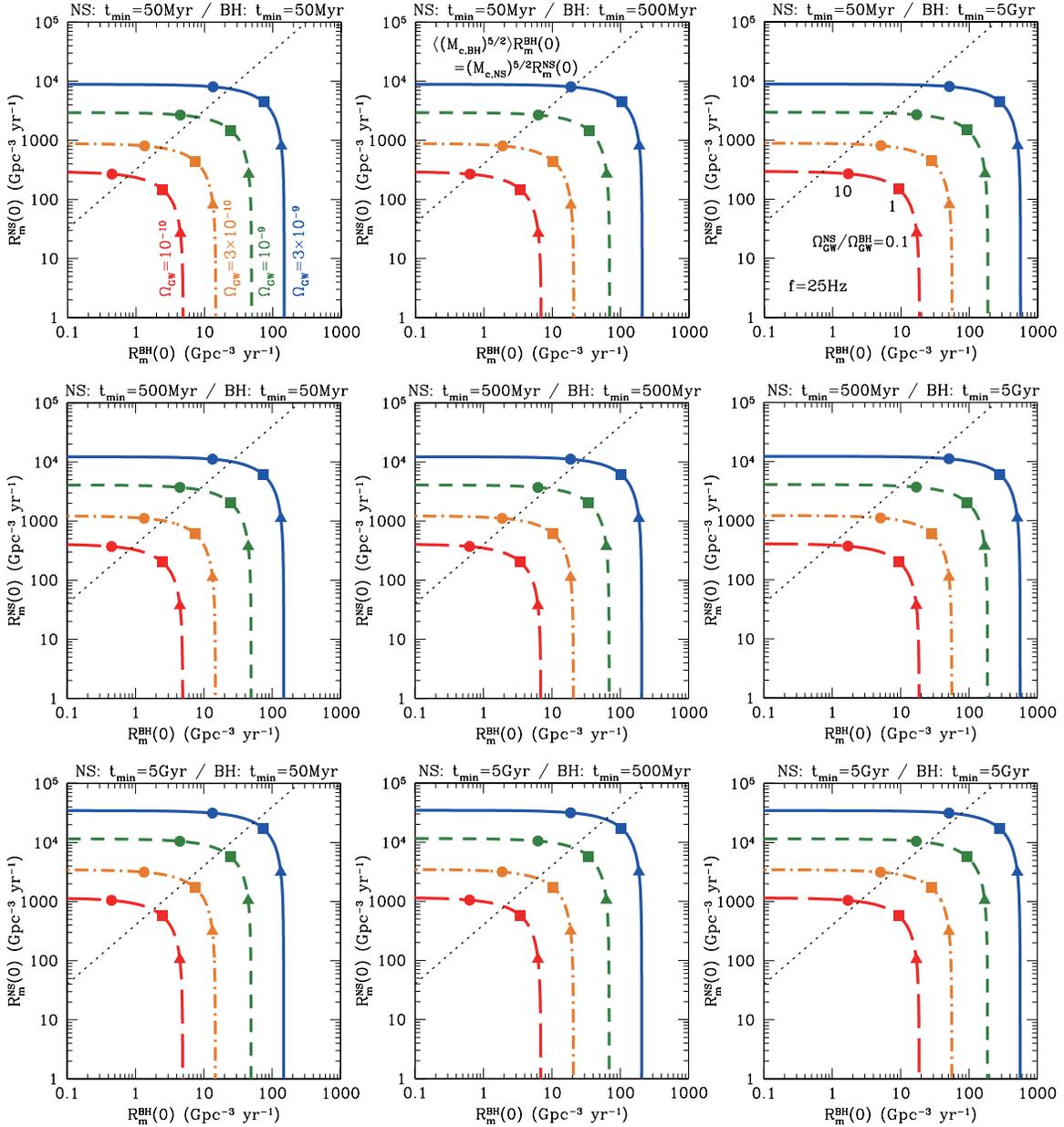}
\epsscale{1.0}
\caption{Total energy density of GW background at $f=25$~Hz plotted on $R^{\rm BH}_m(0)$ vs $R^{\rm NS}_m(0)$ plane for various values of minimum delay time, $t_{\rm min}$. Thick lines correspond, from top to bottom, to $\Omega_{\rm GW}(25~{\rm Hz})=3\times10^{-9}$, $10^{-9}$, $3\times10^{-10}$ and $10^{-10}$. Circles, squares and triangles represent the points with $\Omega^{\rm NS}_{\rm GW}/\Omega^{\rm BH}_{\rm GW}=10$, 1 and 0.1, respectively. Thin dotted lines denote $\langle (M_{c,{\rm BH}})^{5/2}\rangle R^{\rm BH}_m(0)=(M_{c,{\rm NS}})^{5/2}R^{\rm NS}_m(0)$.}
\label{fig:map}
\end{figure*}

In Figure~\ref{fig:intro}, we compare the GW background spectrum computed with the model proposed by \citet{ajith11} and the approximation of Equation~(\ref{eq:gwbgispr}). Since the estimated range of the local merger rate is $R^{\rm BH}_m(0)=9$--240~Gpc$^{-3}$~yr$^{-1}$ \citep[][]{LIGOf} and the investigated range of the minimum delay time in this study is $t_{\rm min} = 50$~Myr--5~Gyr, we show spectra in the cases with $(R^{\rm BH}_m(0),t_{\rm min})=(240~{\rm Gpc}^{-3}{\rm yr}^{-1},50~{\rm Myr})$ for the maximum and $(R^{\rm BH}_m(0),t_{\rm min})=(9~{\rm Gpc}^{-3}{\rm yr}^{-1},5~{\rm Gyr})$ for the minimum. For both models, we find that the difference in energy density of GW background between the model from \citet{ajith11} and approximation (\ref{eq:gwbgispr}) is at most 15\% in the frequency range of $f < 150$~Hz. According to the statistical study of \citet{call16}, Advanced LIGO can hardly distinguish the spectral difference from a simple $f^{2/3}$ power law for the GW background. As shown in Figure~\ref{fig:intro}, compared to the case assuming that the all binary BHs have the same masses as in GW150914 \citep[e.g., fiducial model of][]{LIGOc}, the GW background spectrum has a lower energy density and additional power at high frequencies due to low-mass BHs. It is consistent with the result of \citet{dvor16}, who calculated the mass distribution of binary BHs based on the initial mass function of the progenitor stars and the relation between the initial mass and BH mass. Incidentally, while the spectrum based on \citet{ajith11} shown in Figure~\ref{fig:intro} has a lower energy than the approximation of Equation~(\ref{eq:gwbgispr}) in this range, it can be either higher or lower depending on the BH spin. The expected sensitivity curves of the network of Advanced LIGO and VIRGO \citep[][]{LIGOc} are also shown in Figure~\ref{fig:intro}. If the event rate of binary BH mergers is found to be as high as $\sim$100~Gpc$^{-3}$~yr$^{-1}$ through the direct detection of the GW signal, the GW background may also be observed even for a case with a long delay time.

Binary NS mergers also contribute to the GW background. In Figure~\ref{fig:map}, we plot the total energy density of the GW background at $f=25$~Hz, $\Omega_{\rm GW}(25~{\rm Hz})=\Omega^{\rm BH}_{\rm GW}(25~{\rm Hz})+\Omega^{\rm NS}_{\rm GW}(25~{\rm Hz})$, on the $R^{\rm BH}_m(0)$ vs $R^{\rm NS}_m(0)$ plane for various values of the minimum delay time, $t_{\rm min}$. Using Equations~(\ref{eq:gwbgispr}) and (\ref{eq:gwbgisprns}), we adopt the mass distribution from the results of the first observational run of the Advanced LIGO detectors \citep[][]{LIGOf} for the chirp mass of BH binaries again and we set the chirp mass of NS binaries to $M_{c,{\rm NS}}=1.2M_\odot$, which corresponds to an equal-mass binary with masses $1.4M_\odot+1.4M_\odot$. Since the energy density of the GW background is proportional to the chirp mass to the power $5/3$, BH binaries may be the dominant sources of the GW background in spite of their lower merger rate. According to a recent theoretical estimation, the local merger rate of binary NSs is $R^{\rm NS}_m(0)=52$--162~Gpc$^{-3}$yr$^{-1}$ and their detection rate for a network of advanced GW detectors is expected to be several events per year \citep[][]{domi15}. In contrast, the NS merger rate in our Galaxy has been estimated to be 21~Myr$^{-1}$ from the observation of a double pulsar system \citep[][]{kim15}, which corresponds to $R^{\rm NS}_m(0)=250$~Gpc$^{-3}$yr$^{-1}$. \citet{hotoke15} evaluated the event rate of the r-process sources to be $\le$90~Myr$^{-1}$ in our Galaxy. If the sites of the r-process elements are NS mergers, the local merger rate will be $R^{\rm NS}_m(0) \le 1000$~Gpc$^{-3}$yr$^{-1}$.

Roughly speaking, the direct GW detection rate is proportional to the chirp mass to the power $5/2$. We draw lines denoting $(M_{c,{\rm BH}})^{5/2}R^{\rm BH}_m(0)=(M_{c,{\rm NS}})^{5/2}R^{\rm NS}_m(0)$ in Figure~\ref{fig:map}. Then, below these lines, the event rate of binary BH merger is higher than that of binary NS merger. In Figure~\ref{fig:map}, we also plot the points where the ratios of contributions from binary NSs and binary BHs to the GW background are $\Omega^{\rm NS}_{\rm GW}/\Omega^{\rm BH}_{\rm GW}=10$, 1 and 0.1. We can see the existence of a region where the binary NS merger has a lower event rate for the direct GW detection but a larger contribution to the GW background than the binary BH merger.

Anyway, GW astronomy will enable us to verify the consistency between the local merger rate and the energy density of the GW background in the near future. Nevertheless, there are other possible sources of the GW background. There could be binary systems of heavy BHs and NSs \citep[][]{domi15,belcz16}, which have not been observed yet. It is possible to apply our study to NS-BH binaries; Equations~(\ref{eq:gwbgispr}) and (\ref{eq:f25bh}) hold for NS-BH binaries because their formation rate is proportional to that of heavy BHs, i.e., low-metallicity stars. The local merger rates of NS-NS, BH-BH and NS-BH binaries can be measured individually by direct GW detection, and their integrated spectrum can be observed as the GW background. In contrast, the metallicity dependence of the initial mass function is beyond the scope of this study. In particular, the first stars (Population III stars) are thought to have a top-heavy initial mass function due to the absence of metals, while whether their contribution to the GW background is efficient \citep[][]{inay16} or negligible \citep[][]{hart16,dvor16} is still an open question. We hope that this study will provide a first step toward understanding the GW background on the basis of the metallicity evolution of galaxies.

\acknowledgments

This work was partially supported by Grants-in-Aids for the Scientific Research (No.~24105008, No.~26104006 and No.~26870615) from MEXT in Japan and University Research Support Grant from the National Astronomical Observatory of Japan (NAOJ).

\end{document}